\documentclass[a4paper]{mem}
\usepackage{natbib}
\usepackage{graphicx}
\usepackage[a4paper]{hyperref}
\begin{document}

\title{How do brown dwarves form?}
\author{A. P. Whitworth, S. P. Goodwin
\thanks{email: ant@astro.cf.ac.uk}}
\institute{School of Physics \& Astronomy, Cardiff University, 5 The Parade, \\
Cardiff CF24 3YB, Wales, UK} 

\abstract{We review and evaluate four mechanisms for forming brown dwarves: 
(i) dynamical ejection of a stellar embryo from its placental prestellar core; 
(ii) opacity-limited fragmentation of a shock-compressed layer; (iii) 
gravitational instabilities in discs, triggered by impulsive interactions 
with other discs or naked stars; and (iv) photo-erosion of pre-existing cores. 
All these mechanisms can produce free-floating brown dwarves, but only (ii) 
and (iii) are likely to produce brown dwarves in multiple systems, and (i) 
has difficulty delivering brown dwarves with discs.
\keywords{Star Formation - Brown Dwarves}}
\authorrunning{A. P. Whitworth \& S. P. Goodwin}
\titlerunning{How do brown dwarves form?}
\maketitle

\section{Introduction}

The existence of brown dwarves was first proposed on theoretical grounds by 
Kumar (1963a,b) and by Hayashi \& Nakano (1963). However, more than three 
decades passed before brown dwarves were observed unambiguously (Rebolo et 
al., 1995; Nakajima et al., 1995; Oppenheimer et al., 1995). Brown dwarves 
are now observed routinely (McCaughrean et al., 1995; Luhman et al. 1998; 
Wilking et al. 1999; Luhman \& Rieke, 1999; Lucas \& Roche, 2000; Mart\'in 
et al., 2000; Luhman et al., 2000; B\'ejar et al., 2001; Mart\'in et al., 
2001; Wilking et al., 2002; McCaughrean et al., 2002; etc.), and so it is 
appropriate to ask how such low-mass objects are formed. In particular, 
astronomers are concerned with the question of whether brown dwarves form 
in the same way as more massive stars, and whether there is a dividing 
line between the mechanisms that produce stars and those that produce 
planetary-mass objects.

This paper is concerned with four possible mechanisms for forming brown 
dwarves. Section \ref{S:EJECTION} considers the possibility that brown 
dwarves are formed when low-mass protostellar embryos are ejected from 
their placental prestellar cores before they can accrete sufficient mass 
to ignite hydrogen (Reipurth \& Clarke, 2001). Section \ref{S:OPACITY} 
considers the possibility that brown dwarves are formed by 
opacity-limited fragmentation in turbulent molecular clouds. Section 
\ref{S:DISC} considers the possibility that brown dwarves are formed by 
gravitational instabilities in circumstellar discs, in particular 
circumstellar discs which are subject to impulsive perturbations due to 
interactions with other discs or with naked stars. Section \ref{S:PHOTO} 
considers the possibility that brown dwarves are formed by the 
photo-erosion of more massive cores which find themselves overrun by HII 
regions (Hester. 1997). Section \ref{S:CONC} summarizes our main conclusions

\section{Ejection} \label{S:EJECTION}

The collapse and fragmentation of a prestellar core is 
unlikely to lead to the formation of a single star. Even 
quite modest levels of turbulence (e.g. Goodwin, 
Whitworth \& Ward-Thompson 
2004a) and/or global rotation (e.g. Cha \& Whitworth, 
2003; Hennebelle et al. 2003, 2004) are sufficient to ensure 
the formation of a small-$N$ cluster of protostars, which 
then grow by competitive accretion and interact dynamically 
(Whitworth et al., 1995; Bonnell et al., 2001). 
Protostars that get ejected from the core before they 
have had time to grow to $0.08\,{\rm M}_\odot$, end 
up as brown dwarves (Reipurth \& Clarke, 2001). It seems 
inescapable that this mechanism occurs, since all that 
is required is the formation and coexistence of more 
than two protostars in a core, with one of them being 
less massive than $0.08\,{\rm M}_\odot$; $N$-body 
dynamics will then almost inevitably eject one of the 
stars, and usually the least massive one. 

Several numerical simulations have been performed, 
using SPH with sink particles, to demonstrate this 
mechanism at work, both in cores with very high 
levels of turbulence (Bate, Bonnell \& Bromm, 2002; 
Delgado-Donate, Clarke \& Bate, 2003, 2004), and in 
cores with more modest levels of turbulence 
(Goodwin, Whitworth \& Ward-Thompson, 2004a,b,c).

The main concern with these simulations is that, by 
invoking sink particles, protostellar embryos are 
instantaneously converted into point masses. This predisposes 
them to dynamical ejection, and prohibits them from 
merging or fragmenting further. Therefore the efficiency 
of the mechanism may have been overestimated, although 
probably not by much.

Additional support for the mechanism comes from 
the recent paper by Goodwin et al. (2004b), which 
presents an ensemble of simulations of the collapse 
and fragmentation of 
cores having a mass spectrum, density profiles, and 
low levels of turbulence, matched to those observed 
in Taurus. These simulations reproduce rather well 
the unusual stellar mass function observed in Taurus 
(Luhman et al., 2003), including the relative paucity 
of brown dwarves. As far as we are aware, these are 
the first simulations to demonstrate a direct causal 
connection between the core mass spectrum and the 
stellar initial mass function.

However, ejection is unlikely to be 
the only mechanism forming brown dwarves, since it 
seems very unlikely to produce brown dwarves in 
multiple systems. It is also unclear 
whether ejected brown dwarves can retain the 
discs which seem to be needed to 
explain the significant fraction of brown dwarves 
having IR excesses and other signatures of on-going 
accretion.

\begin{figure*}
\centering
\includegraphics[angle=-90,width=11.0cm]{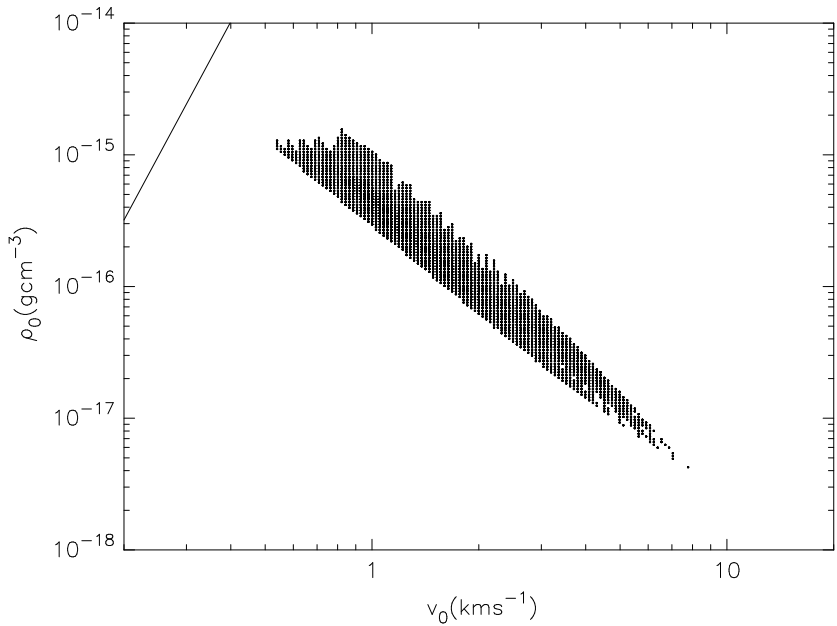}
\caption{A log/log plot of the $(\rho,v)$ plane. The dots mark 
combinations of pre-shock density. $\rho$, and collision speed, 
$v$, for which the fastest growing fragment has a mass less than 
$0.005\,{\rm M}_\odot$; we assume that the effective post-shock 
sound speed is $\sigma = 0.2\,{\rm km}\,{\rm s}^{-1}$, 
corresponding to molecular gas at $10\,{\rm K}$. The 
irregularities in the boundaries of this region reflect the 
tendency of low-mass fragments to undergo pulsations before 
they collapse. The solid line is the locus below which $\rho$ 
must fall if our treatment of the radiation from the accretion 
shock is to be valid; see Boyd \& Whitworth (2004) for details.}
\label{FIG:OPACITY}
\end{figure*}

\section{Opacity-limited fragmentation} \label{S:OPACITY}

Conventionally, the minimum mass for star formation 
has been evaluated on the basis of the 3D hierarchical 
fragmentation picture developed by Hoyle (1953). In 
this picture, a large protocluster cloud becomes Jeans 
unstable and starts to contract. As long as the sound 
speed in the gas remains approximately constant, the 
increasing density reduces the Jeans mass, and 
eventually separate parts of the cloud (subclouds) 
become Jeans unstable and can contract independently 
of one another. This process repeats itself recursively, 
breaking the cloud up into ever smaller and denser 
subsub...subclouds, until the gas becomes so opaque 
that it can no longer radiate away the gravitational 
energy being released by contraction. At this stage 
the gas starts to heat up, and fragmentation ceases. 
This yields a minimum mass in the range $M_{_{\rm MIN}} 
\sim 0.007\,{\rm M}_\odot$ to $M_{_{\rm MIN}} \sim 
0.015\,{\rm M}_\odot\,$ (e.g. Rees, 1976; Low \& 
Lynden-Bell, 1976; Silk, 1977).

However, it appears that 3D hierarchical fragmentation 
does not work. There is no evidence of its occuring 
in nature, nor does it occur in numerical simulations 
of star formation. Therefore one must question 
estimates of $M_{_{\rm MIN}}$ 
based on 3D hierarchical fragmentation. The reason 
3D hierarchical fragmentation does not work probably 
has to do with the fact the timescale on which a 
fragment condenses out in 3D is always longer than 
the timescale on which the parent cloud (of which it 
is part) is contracting. Therefore fragmentation, if 
it occurs at all, is only temporary, and the fragments 
are then merged by the overall contraction of the parent 
cloud. The only way to avoid this is to start with 
proto-fragments which are widely spaced, but then 
the rate of accretion onto a fragment is very high, 
and even if it starts off with mass $M_{_{\rm FRAG}} 
\sim M_{_{\rm JEANS}}\,$, it will have grown much more 
massive by the time its contraction becomes non-linear.

We have therefore revisited the question of the minimum 
mass for star formation, but now using a model which 
invokes 2D one-shot fragmentation of a shock-compressed 
layer. We argue that this model is more relevant to the 
contemporary picture of `star formation in a crossing 
time' (Elmegreen, 2000). In this picture star formation 
occurs in molecular clouds wherever two -- or more -- 
turbulent flows of sufficient density collide with 
sufficient ram pressure to produce a shock-compressed 
layer out of which prestellar cores can condense. This 
model is 2D because fragmentation of a shock-compressed 
layer is in effect two-dimensional (the motions which 
initially assemble a fragment are in the 
plane of the layer), and it is `one-shot' in the sense 
of not being hierarchical or recursive.

A shock compressed layer is contained by the ram 
pressure of the inflowing gas, and until it fragments 
it has a rather flat density profile. Normally it 
fragments whilst it is still accumulating, at time 
$t_{_{\rm FRAG}}$, and the fastest growing fragment 
has mass $m_{_{\rm FRAG}}$, radius $r_{_{\rm FRAG}}$ 
(in the plane of the layer) and half-thickness 
$z_{_{\rm FRAG}}$ (perpendicular to the plane of the 
layer) given by
\begin{eqnarray}
t_{_{\rm FRAG}} & = & \left( \sigma / G\,\rho\,v \right)^{1/2} \,, \\ 
m_{_{\rm FRAG}} & = & \left( \sigma^7 / G^3\,\rho\,v \right)^{1/2} \,, \\
r_{_{\rm FRAG}} & = & \left( \sigma^3 / G\,\rho\,v \right)^{1/2} \,, \\
z_{_{\rm FRAG}} & = & \left( \sigma^5 / G\,\rho\,v^3 \right)^{1/2} \,,
\end{eqnarray}
where $\sigma$ is the net velocity dispersion in the 
shock-compressed layer, $\rho$ is the pre-shock density 
in the colliding flows, and $v$ is the relative speed 
with which the flows collide. We note (a) that the 
fragments are initially flattened objects ($r_{_{\rm FRAG}} 
/ z_{_{\rm FRAG}} \sim v / \sigma \gg 1$), and (b) that 
$m_{_{\rm FRAG}}$ is not simply the 3D Jeans mass evaluated 
at the post-shock density and velocity dispersion -- it is 
larger by a factor $(v/\sigma)^{1/2}$. 

2D one-shot fragmentation has the advantage that the 
fastest-condensing fragment has finite size, i.e. fragments 
with initial radius $\sim r_{_{\rm FRAG}}$ condense out 
faster than either larger or smaller fragments. Moreover 
we can analyze the growth of a fragment in a shock-compressed 
layer, taking account of the continuing inflow of matter into 
the fragment. Hence we can identify the smallest 
fragment which can cool radiatively fast enough to dispose 
of {\it both} the $PdV$ work being done by compression of the 
fragment, {\it and} the energy being dissipated at the accretion 
shock where matter continues to flow into the fragment; these 
two sources of heat turn out to be comparable. We find 
(Boyd \& Whitworth, 2004) that 
for shocked gas with temperature $T \sim 10\,{\rm K}$ and 
no turbulence (i.e. velocity dispersion equal to the 
isothermal sound speed, $0.2\,{\rm km}\,{\rm s}^{-1}$), 
the smallest fragment which can condense out is less than 
$0.003\,{\rm M}_\odot$, and fragments with mass below 
$0.005\,{\rm M}_\odot$ condense out for a wide range of 
pre-shock density $\rho$ and shock speed $v$ (as 
illustrated on Figure \ref{FIG:OPACITY}). We emphasize that this 
analysis is more robust than the standard one based on 
3D hierarchical fragmentation, on two counts. (i) The 
fragments have condensation timescales shorter than all 
competing length-scales (a well-known property of layer 
fragmentation), so they do not tend to get merged by the 
overall contraction of a parent fragment. (ii) Ongoing 
accretion is taken into 
account; indeed the smallest fragment of all starts off 
with mass $0.0011\,{\rm M}_\odot$ and grows to 
$0.0027\,{\rm M}_\odot$ before its contraction becomes 
non-linear. We conclude (Boyd \& Whitworth, 2004) that 
brown dwarves and planetary-mass objects with masses 
down to $0.003\,{\rm M}_\odot$ can condense out of 
shock-compressed layers, along with more massive stars. 

\section{Disc instabilities} \label{S:DISC}

Brown dwarves may also form via gravitational 
instabilities in massive protostellar discs. 
If we consider a massive disc in isolation, there is some 
doubt as to whether it will fragment gravitationally, 
spawning low-mass companions to the central primary 
protostar, or whether spiral instabilities will act 
to quickly redistribute angular momentum, thereby 
stabilizing -- and ultimately disspating -- the disc 
before it can fragment. However, if a massive 
protostellar disc interacts impulsvely with another 
disc, or with a naked star, then it can be launched 
directly into the non-linear regime of gravitational 
instability and fragmentation is then much more likely. 
Such interactions must be quite frequent in the dense 
proto-cluster environment where stars are born; for 
example, young massive protostellar discs extend out 
to several hundred AU, and $40\%$ of stars are born 
in binary systems with semi-major axes less than 
$100\,{\rm AU}$. Boffin et al. (1998) and Watkins et 
al. (1998a,b) have simulated parabolic interactions 
between two protostellar discs, and between a single 
protostellar disc and a naked protostar. The protostars 
all have mass $M_{_\star} = {\rm M}_\odot\,$, and 
the discs also have $M_{_{\rm DISC}} = {\rm M}_\odot\,$ 
(so these are very young protostars with very massive discs). 
All possible mutual orientations of spin and orbit are 
sampled. The critical parameter turns out to be the 
effective shear viscosity in the disc. If the 
Shakura-Sunyaev parameter is low, $\alpha_{_{\rm SS}} 
\sim 10^{-3}\,$, the interactions produce mainly 
planetary-mass companions, i.e. objects in the range 
$\sim 0.001\,{\rm M}_\odot$ to $\sim 0.01\,{\rm M}_\odot\,$. 
Conversely, if $\alpha_{_{\rm SS}}$ is larger, 
$\alpha_{_{\rm SS}} \sim 10^{-2}\,$, interactions 
produce mainly brown-dwarf companions, i.e. objects 
in the range $\sim 0.01\,{\rm M}_\odot$ to 
$\sim 0.1\,{\rm M}_\odot\,$. The formation of low-mass 
companions is most efficient for interactions in which 
the orbital and spin angular momenta are all parallel; 
on average 2.4 companions are formed per interaction in 
this case. If the orbital and spin angular momenta are 
randomly oriented with respect to each other, then on 
average 1.2 companions are formed per interaction. This 
is evidently a good way of producing brown dwarves and 
planetary-mass objects as companion objects. It can also produce 
brown dwarves with discs.

\section{Photo-erosion of pre-existing prestellar cores} \label{S:PHOTO}

Another mechanism for producing brown dwarves is to 
start with a standard prestellar core (one which if 
left to its own devices is destined to form an 
intermediate- or high-mass star), and have it overrun 
by an HII region (Hester, 1996). As a result, an ionization front (IF) 
starts to eat into the core, `photo-eroding' it. At 
the same time, a compression wave (CW) advances into 
the core ahead of the IF. When the CW reaches the 
centre, a protostar is created, which then grows by 
accretion. At the same time, an expansion wave (EW) 
is reflected and propagates outwards, setting up the 
inflow which feeds accretion onto the central protostar. 
The outward propagating EW soon meets the inward 
propagating IF, and shortly thereafter the IF finds 
itself ionizing gas which is so tightly bound to the 
protostar that it cannot be unbound by the act of 
ionization. All the material interior to the IF at 
this juncture ends up in the protostar. On the basis 
of a simple semi-analytic treatment, Whitworth \& 
Zinnecker (2004) show that the final mass is given by
\begin{eqnarray} \nonumber
M & \simeq & 0.01\,{\rm M}_\odot\,
\left( \frac{a_{_{\rm I}}}{0.3\,{\rm km}\,{\rm s}^{-1}} 
\right)^6 \,\times \\ \nonumber
 & & \left( \frac{\dot{\cal N}_{_{\rm LyC}}}{10^{50}\,
{\rm s}^{-1}} \right)^{-1/3} \,\left( \frac{n_{_0}}{10^3\,
{\rm cm}^{-3}} \right)^{-1/3} \,,
\end{eqnarray}
where $a_{_{\rm I}}$ is the isothermal sound speed 
in the neutral gas of the core, $\dot{\cal N}_{_{\rm LyC}}$ 
is the rate at which the star(s) exciting the HII region 
emit hydrogen-ionizing photons, and $n_{_0}$ is the density 
in the ambient HII region. 

The mechanism is rather effective, in the sense that 
it produces brown dwarves for a wide range of conditions. 
Indeed, the EGGs identified in M16 by Hester et al. 
(1996) would appear to be pre-existing cores being photoeroded 
in the manner we describe. However, the mechanism is also 
very inefficient, in the sense that it usually takes a rather 
massive initial prestellar core to form a single very 
low-mass brown dwarf or planetary-mass object. Moreover, 
the mechanism can only work in the immediate vicinity of 
an OB star, so it cannot explain the formation of all 
brown dwarves, only free-floating brown dwarves in HII regions. 
Brown dwarves formed in this way are likely to be single. 
They should have no difficulty retaining small discs. 

\section{Conclusions} \label{S:CONC}

Four mechanisms for forming brown dwarves have been 
described. The question of which, if any, of these 
mechanisms contributes to brown dwarf formation in nature 
may be settled once the binary statistics of brown 
dwarves are known accurately, and the frequency of accretion 
discs around brown dwarves is established. If any of these 
four mechanisms are important, this suggests that the 
formation of brown dwarves forms a continuum with the 
formation of more massive stars.

\begin{acknowledgements}
We gratefully acknowledge the support of a PPARC Research 
Associateship (Ref. PPA/G/S/1998/00623) and an EC Research 
Training Network awarded under the Fifth Framework  (Ref. 
HPRN-CT-2000-00155).
\end{acknowledgements}

\bibliographystyle{aa}

\end{document}